\documentclass[useAMS,usegraphicx]{mn2e}

\title[X-ray absorption by BLR clouds in Mrk 766]{X-ray absorption by Broad Line Region Clouds in Mrk 766}
\author[G. Risaliti et al.]
{G. Risaliti,$^{1,2}$ 
E. Nardini,$^{3}$
M.~Salvati,$^2$ M.~Elvis,$^1$ 
G.~Fabbiano,$^1$  R.~Maiolino,$^4$
\newauthor P. Pietrini,$^{3}$ G. Torricelli-Ciamponi$^2$\\
$^1$ Harvard-Smithsonian Center for Astrophysics, 60 Garden St. 
Cambridge, MA 02138 USA {E-mail: grisaliti@cfa.harvard.edu}\\
$^2$ INAF - Osservatorio Astrofisico di Arcetri, L.go E. Fermi 5,
Firenze, Italy\\
$^3$ Universit\`a di Firenze, Largo E.~Fermi 2, Firenze, Italy\\
$^4$ INAF - Osservatorio Astronomico di Roma, via di Frascati 22, Monte Porzio Catone, 00040, Italy
}
\begin{document}

\date{Released Xxxx Xxxxx XX}

\pagerange{\pageref{firstpage}--\pageref{lastpage}} \pubyear{2002}

\maketitle

\label{firstpage}

\begin{abstract}
We present a new analysis of a 9-day long XMM-Newton monitoring of the Narrow Line Seyfert~1 galaxy Mrk~766.
{ We show that the strong changes in spectral shape which occurred during this observation can be interpreted as due to } 
Broad Line Region clouds crossing the line of sight to the X-ray source. { Within the occultation scenario}, the spectral and temporal analysis of the eclipses provides precise estimates of the geometrical structure, location and physical properties of the absorbing clouds. In particular, we show that these clouds have cores with column densities of at least a few 10$^{23}$~cm$^{-2}$ and velocities in the plane of the sky of the order of thousands km/s. The three different eclipses monitored by XMM-Newton suggest a broad range in cloud velocities (by a factor $\sim$4-5). Moreover, two iron absorption lines clearly associated with each eclipse suggest the presence of highly ionized gas around the obscuring clouds, and an outflow component of the velocity spanning from 3,000 to 15,000~km/s.
\end{abstract}


\section{Introduction}
Time variability of X-ray absorption is a common feature in Active Galactive Nuclei (AGN).
An analysis of column density ($N_H$) variations, performed on a sample of bright obscured AGN with
multiple hard X-ray (1-10~keV) observations (Risaliti et al.~2002) revealed that $N_H$ variability is almost
ubiquitous among nearby Seyfert galaxies on time scales from a few months to a few years.
In order to investigate shorter variability time scales, two strategies are possible: dedicated campaigns of multiple observations, or time-resolved spectral analysis of single, long observations.
In the past few years, both methods have been used to discover $N_H$ variations on time scales from 
a few days down to a few hours for a handful of sources: NGC~1365 (Risaliti et al.~2005, 2007, 2009), NGC~4388 (Elvis et al.~2004), NGC~4151 (Puccetti et al.~2007), NGC~7582 (Bianchi et al.~2009).
The physical implications of these results are that the X-ray absorber must be made of clouds with linear dimensions of the order of 10$^{13}$-10$^{14}$~cm, velocities in excess of 10$^3$~km~s$^{-1}$, and densities n$\sim$10$^{10}$-10$^{11}$~cm$^{-3}$. Assuming that these clouds are orbiting around the central black hole with Keplerian velocities, the inferred distances from the X-ray sources are of the order of thousands of gravitational radii (R$_G$), corresponding to 
10$^{16}$-10$^{17}$~cm for black hole masses in the range from 10$^6$ to a few 10$^7$ solar masses.  
These physical parameters are typical of the clouds emitting the broad lines observed in the AGN optical and UV spectra. It is therefore natural to conclude that the
variable X-ray absorber and the Broad Line Region (BLR) clouds are one and the same. 

In order to further test this scenario in other sources, we started a systematic analysis of all the
archival long ($>$100~ks) observations of bright AGN (intrinsic 2-10~keV of at least 
10$^{-11}$~erg~s$^{-1}$~cm$^{-2}$) in order to search for occultations due to obscuring clouds crossing
the line of sight to the X-ray source. Our approach is based on a preliminary analysis of the 
hardness-ratio (HR) light curve, with the aim of selecting the time intervals with the strongest spectral
variations, possibly due to occultations, and a subsequent complete spectral analysis of the 
selected intervals. 
A detailed description of this method is illustrated in Risaliti et al.~2009A, 2009B 
for two {\em XMM-Newton} observations of NGC~1365. 

Here we present a new analysis of a long ($\sim$800~ks) {\em XMM-Newton} observation of the AGN in Mrk~766 (z=0.0129), 
which exhibits strong signatures of multiple occultations.

The Narrow Line Seyfert~1 (NLS1)  galaxy Mrk~766 is highly variable in the hard X-ray (2-10~keV) flux, 
on time scales as short as
a few hundred seconds, as typical for NLS1s (e.g. McHardy et al.~2004). 
{ Being one of the X-ray brightest sources of its class, it has been observed intensively with the
main X-ray observatories, such as ROSAT (Molendi \& Maccacaro~1994), ASCA (Leighly et al.~1996, Nandra et al.~1997), {\em BeppoSAX} (Matt et al. 2000). More recently, higher-quality {\em XMM-Newton} observations allowed detailed studies of the X-ray variability (Page et al.~2001, Boller et al.~2001, Vaughan \& Fabian~2003), and revealed the presence of absorption due to highly ionized iron lines (Pounds et al.~2003).} 

The longest and highest quality X-ray observation available for Mrk~766 has been performed with {\em XMM-Newton} in 2006 over six consecutive orbits, resulting in a total elapsed time of about 10 days of quasi-continuous monitoring. 
A complete time-resolved spectral analysis of these data has been presented by Miller et al.~(2007) and Turner et al.~(2007, hereafter T07). In these works, the observation is divided in time intervals of 25~ks, and for each interval a detailed spectral analysis is performed, testing two different scenarios: in the first one, the observed spectral and flux variability is ascribed to a scattering component; in the second one, the dominant changing component is a warm absorber partially covering the central source. Both scenarios can adequately reproduce the observed spectral and flux variability. Interestingly, in both models an extra component is needed, consisting of a partial covering by a relatively low-ionization cloud with column density in the range 10$^{22}$-10$^{23}$~cm$^{-2}$.
{  The best fit values of the 
column density and the covering factor of this component, tabulated in T07, follow a similar variability pattern in both the scattering and absorption scenarios. In particular, in both models these parameters have the highest values, and the  highest statistical significance, during the first orbit and the first part of the second orbit, while they are only marginally significant for the remainder of the observation.
The similar behaviour of N$_H$(t) in both scenarios suggests that it is possible to study this component in detail independently of the interpretation adopted for the underlying spectrum.  }

Here we concentrate our analysis on this varying cold absorbtion component, 
trying to isolate its effects from those of the other components. 
We will show that after an appropriate selection  of the time intervals the spectral analysis { strongly suggests}
the presence of three occultation events, due to three different absorbers, with physical parameters typical of BLR clouds. 

\begin{figure*}
\includegraphics[width=17.0cm,angle=0]{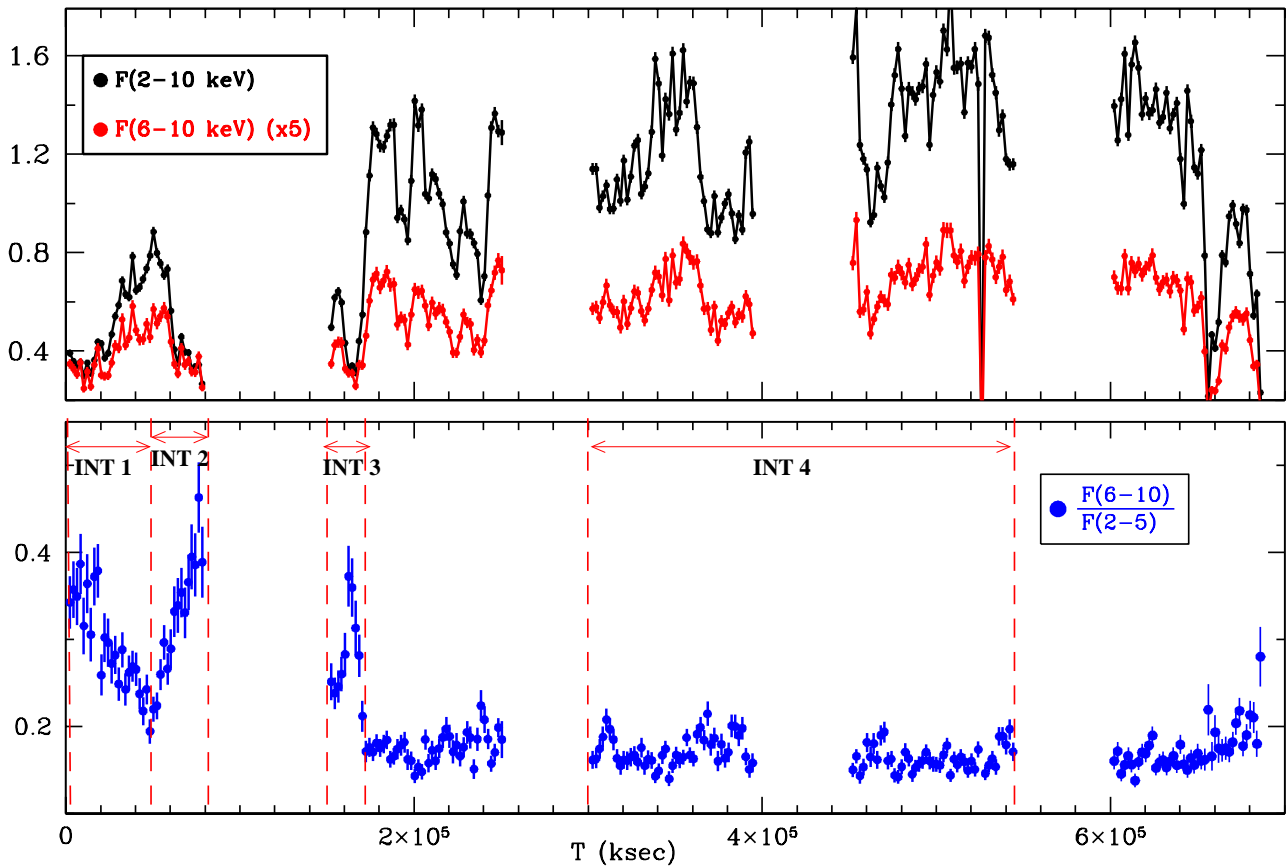}
\caption{Flux (top, in arbitrary units) and hardness ratio HR=F(6-10~keV)/F(2-5~keV) (bottom)
light curves of the {\em XMM-Newton} observation of Mrk~766. For clarity, only EPIC-PN data are shown.
}
\label{totfit}
\end{figure*}

\section{Reduction and data analysis.}
Mrk~766 was observed by {\em XMM-Newton} in six consecutive orbits from May 23 to June 03, 2005 (the observation in the sixth orbit is short ($\sim$30~ks) and is not analyzed here).
The data have been reduced and analyzed following a standard procedure, which we already described in other similar works (e.g. Risaliti et al.~2009). The only non-completely straightforward aspect of the reduction is the
choice of the rejection level for the background flares. Since the source is bright (F(2-10~keV)$\sim$2$\times$10$^{-11}$~erg~cm$^{-2}$s$^{-1}$), we can use a higher threshold than the conservative level suggested in the
{\em XMM-Newton} reduction guide. In particular, since we are interested in a time-resolved study, 
we can accept a relatively high noise in some time intervals, which otherwise would be completely neglected in the analysis. We therefore removed only the time intervals where the background flares appear to saturate the
instruments. This led to the rejection of less than 20\% of the observing time.
 Both the temporal and spectral analysis were performed using the PN and combined MOS spectra and light curves.
On average, the MOS data are more noisy than the PN one, and their contribution to the total counts is of the order of 30\%. Therefore, all the results are dominated by the contribution of the PN instrument.
 
Our analysis consists in three main steps: (1)  a temporal analysis of the flux and hardness ratio light curves,
 in order to determine the time intervals with significant spectral variations; (2) an analysis of the average spectra of each of these intervals, in order to determine whether the variations are better explained by absorption variability or by changes in the underlying spectral shape, and (3) 
a detailed time-resolved spectral analysis of the events, { under the assumption that N$_H$ variability is indeed the driving phenomenon. }
We restrict our analysis to the data at energies E$>$1.5~keV.
{ We note that  this choice, while optimal to simplify the analysis of the variable component at E$>$2~keV (and therefore, to search for  neutral N$_H$ variations greater than a few 10$^{22}$~cm$^{-2}$), prevents us from obtaining a global view of the possible reflector/warm absorber affecting the soft part of the spectrum. For an analysis of this component, we refer to T07.}   

 Each step of the analysis is described below.
\subsection{Temporal analysis} 
In Fig.~1 we show
the 2-10~keV flux light curve for the whole observation, together with the light curve of the (6-10~keV)/(2-5~keV) hardness ratio. As expected from previous studies (T07), the flux light curve shows strong and rapid variability, on
time scales shorter than one~ks. On average, such strong variations are not observed in the hardness ratio light
curve, suggesting that most of the observed variability consists of flux changes with a $\sim$constant spectral shape. The only evident exceptions are present in the first and second orbit, where we detect strong changes of the hardness ratio, implying variability of the spectral shape. We note that the increase of the hardness ratio in these time intervals is associated to a decrease in the total flux. Therefore, these variations may be due either to (1) a hardening of the intrinsic continuum, with a pivot point in the hard ($>$5~keV) part of the spectrum, so that the soft flux decreases more than the hard one, or (2) to absorption by clouds with column density $N_H$ in the interval $10^{22}$-5$\times$10$^{23}$~cm$^{-2}$, which would cause strong absorption in the 2-5~keV band, but little variation in the
6-10~keV band.
\begin{table}
\caption{Baseline model in the 1.5-10 keV range - 3$^{rd}$ and 4$^{th}$ orbits}
\centerline{\begin{tabular}{lc|lc}
\hline
$\Gamma^a$     & 2.26$^{+0.04}_{-0.04}$  & R$^b$         & 2.2$^{+0.5}_{-0.8}$ \\
E$_1$(Fe)$^c$  & 6.40$^{+0.02}_{-0.02}$  & E$_2$(Fe)$^c$ & 6.67$^{+0.05}_{-0.06}$ \\
W$_1^d$        & $<0.05$                 & W$_2^d$       & 0.14$^{+0.08}_{-0.07}$\\
EW$_1$(Fe)$^e$ & 30$^{+7}_{-8}$          & EW$_2$(Fe)$^e$& 56$^{+24}_{-20}$   \\
F(2-10)$^f$    & 1.39$\times$10$^{-11}$  & L(2-10)$^g$   & 5.2$\times$10$^{42}$ \\
$\chi^2$/d.o.f.& 2419/2362 \\   
\hline
\end{tabular}}
{\footnotesize
$^a$: photon index of the primary continuum; $^b$: reflection efficiency, defined as the ratio between the normalization of the reflection and direct continuum components; $^c$, $^d$, $^e$: peak energy (in keV), width (in keV) and equivalent width (in eV) of the iron emission lines; 
$^f$, $^g$: 2-10~keV flux and luminosity, in c.g.s. units,
adopting a (h$_0$, $\Omega_M$, $\Omega_\lambda$)=(0.7,0.3,0.7) cosmology. All quoted errors, in this table and elsewhere in the paper, are at the 90\% confidence level for one interesting parameter. 
}
\end{table}

\begin{figure}
\includegraphics[width=8.4cm,angle=0]{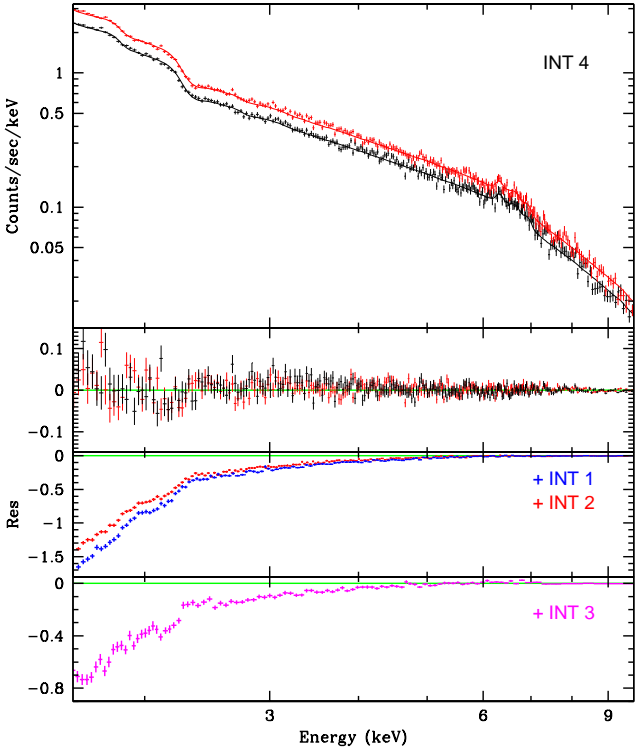}
\caption{Top two panels: data, best fit model and residuals for the spectra obtained from
the third (black marks) and fourth (red marks) orbit, labeled as INT~4 in Fig.~1.  3$^{rd}$ panel: residuals with respect to the best model of INT~4 of the two spectra extracted from the two time intervals INT~1 and INT~2 in the first orbit. Bottom panel: same, for the interval INT~3 in Fig.~1. Note the difference by a factor of $\sim$10 in the residuals scale among the different spectra. For clarity, only data at E$>$2.5~keV are plotted, in order to better show the cutoff-like shape of the residuals.
}
\label{f2}
\end{figure}
\subsection{ Spectral analysis of the time-averaged spectra} 
We extracted a spectrum from the third and fourth orbit (INT~4 in Fig.~1), 
where the emission remains constant in shape above 2~keV,
and determined a model well fitting our data, with reduced $\chi^2$$\sim$1, and no obvious residual features.
This model consists of a power law continuum, two Gaussian emission lines at 6.4~keV and 6.7~keV, respectively, 
and a cold reflection continuum, modeled through the PEXRAV component (Magdziarz \& Zdziarski~1995) in XSPEC. As shown in Fig.~2 and Table~1, this simple model is a good representation of the data. 
{ We note that the reflected continuum is quite high (R$\sim$2, implying a high covering fraction by Compton-thick gas) with respect to the equivalent width of the 
narrow iron emission line (EW$\sim$35~eV). This may be due to either a low iron abundance (a factor of $\sim$2 under-abundance of iron would be enough to reconcile at a 90\% confidence level the R/EW ratio with that expected by George \& Fabian~(1991) for an inclination of 30~degrees), or to an over-simplification of our fit, which reproduces the whole curvature  with respect to a simple power law above $\sim3-4$~keV with this single component, while the actual circumnuclear absorber/reflector may be more complex.
In order to test the consistency of our model, we tried (a) fixing the R/EW ratio to the values predicted by George \ Fabian~(1991), and (b) replacing the PEXRAV + line components with the self-consistent REFLIONX model (Ross \& Fabian~2005). In both cases we obtained a fully acceptable fit ($\chi^2$=2435/2365 d.o.f., and 2433/2364 d.o.f., respectively). }
A more detailed analysis, such as the one presented in T07, can unveil a higher complexity, and the presence of partially covering ionized absorbers. However, we stress that our main aim here is not
a detailed study of these components, but, rather, to obtain a reference model of the time-averaged underlying emission of the source.\\
We then added to the analysis three new spectra, obtained from the first and second half, respectively, of the first orbit (INT~1 and INT~2 in Fig.~1), and from the interval with high hardness ratio in the second orbit (INT~3). We fitted these spectra with the same model as above, plus an extra component consisting of a partial absorber with free column density and covering fraction. A visual analysis 
of the residuals (Fig.~5) suggested to further add two Gaussian components, in order to fit two absorption features in the 6.7-7.5 spectral range.
\begin{figure}
\includegraphics[width=8.5cm,angle=0]{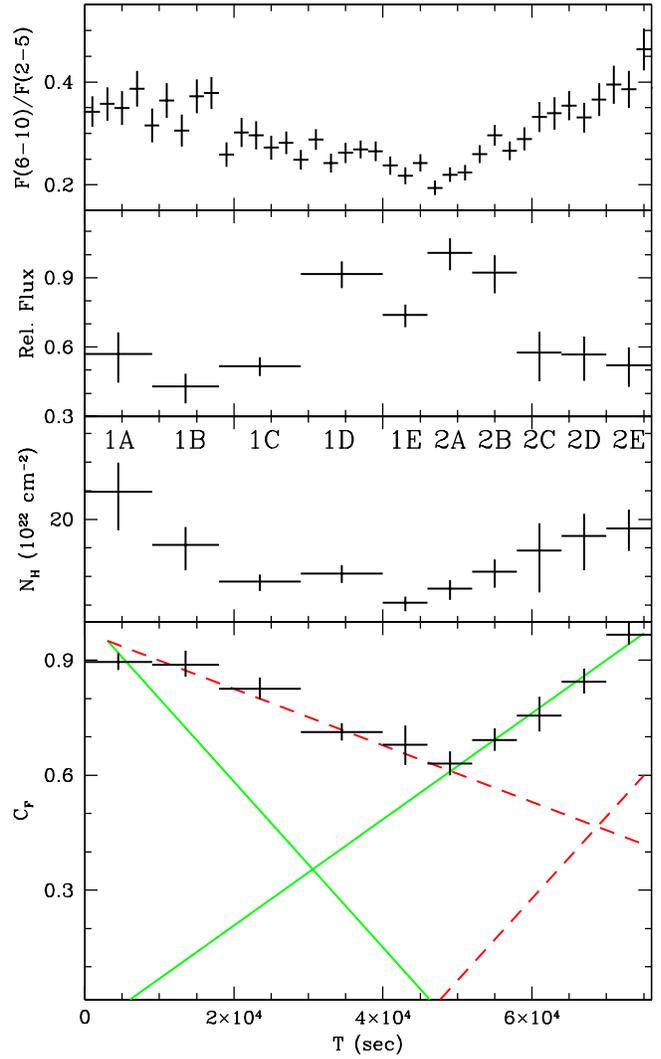}
\caption{Results of the temporal and spectral analysis of the first orbit. { Upper panel:  hardness ratio light curve. This plot is just a zoom of the second panel of Fig.~1. Second panel: intrinsic 2-10~keV flux light curve, normalized to the value measured in INT~4.} Third panel: column density light curve, obtained from the spectral analysis described in the text. Bottom panel: Covering factor (CF)  light curve. The two pairs of lines (continuous/blue and dashed/red) show two possible temporal evolutions of the CF of the two individual clouds giving rise to the observed  curve (see text for details).
}
\label{totfit}
\end{figure}
The fits were done separately for each interval, leaving all the spectral parameters free, only requiring a constant reflection component through the whole observation. 
The main results are the following: \\
(a) In all cases we obtained a good fit ($\chi^2_{red}$$\sim$1), with the partial covering component highly significant in each of the three intervals.\\
(b) The best fit values of the column densities and covering factors of the absorber are different in each interval.\\
(c) The best fit slope of the continuum power law is instead constant within the errors in all the spectra (including the new ones and the reference ones obtained from the third and fourth orbit). { If we repeat the fits without the partial covering absorber (so forcing the continuum slope to adjust to reproduce the spectral variations) we obtain a much worse fit, rejected at a 99.99\% level in all the three fits to INT~1, INT~2 and INT~3. This shows that a change in the continuum slope cannot reproduce the observed spectral variability. Similarly, we also tried to leave the parameters of the reflection component free in each fit, and we found that the results do not change significantly. }  \\
(d) The two absorption features in the 6.7-7.5~keV interval are highly significant (Table~4), and compatible with 
the FeXXV K$\alpha$ and FeXXVI K$\alpha$ absorption lines at rest-frame energies of 6.70 and 6.96~keV, respectively. 
We fitted the two lines fixing their energies and requiring a common shift.
The results (Table~4) indicate that the highly ionized absorber responsible for these features is in outflow, with velocities from $\sim$3,000 to $\sim$15,000~km~s$^{-1}$. The three outflow velocities are not compatible with a constant value. Interestingly, the absorption features are present only when a cold absorber is also present, { while they are not detected in INT~4. Since the integration time of INT~4 is much longer than in the three intervals with evidence of occultations, it may be that a variable highly ionized absorber is present also in INT~4, but its signatures are averaged out in the total spectrum. In order to check this possibility, we divided INT~4 in eight intervals, with a fixed duration of 25~ks,  and repeated the same spectral analysis as for the total INT~4 spectrum. In no case we detected an iron absorption line at a 90\% significance level. The estimated upper limits for each INT~4 sub-interval vary between 20 and~30~eV for the single lines, well below the level detected during the intervals INT~1-3. We therefore conclude that the ionized absorber responsible for the iron absorption lines is physically associated  with the cold absorber.}
\begin{table}
{
\caption{Spectral analysis of the three high-HR intervals}
\centerline{\begin{tabular}{lccccc}
\hline
INT      &  N$_H$ & C$_F$ & F/F$_*$ & $\Gamma$ & $\chi^2$/d.o.f.\\
\hline
1 & 15.0$^{+2.9}_{-2.3}$ & 0.73$^{+0.03}_{-0.03}$ & 0.72$^{+0.18}_{-0.14}$ &2.26$^{+0.09}_{-0.06}$ &  1238/1230 \\
2 & 18.9$^{+4.0}_{-3.1}$ & 0.67$^{+0.04}_{-0.05}$ & 0.74$^{+0.20}_{-0.18}$ &2.27$^{+0.05}_{-0.09}$ &                         993/1027 \\
3 & 18.0$^{+5.2}_{-4.9}$  &  0.62$^{+0.05}_{-0.09}$ & 0.79$^{+0.31}_{-0.22}$ & 2.30$^{+0.12}_{-0.10}$   & 776/803 \\
\hline
\end{tabular}}
\footnotesize{Best fit parameters for the three intervals INT~1, INT~2 and INT~3 (Fig.~1). Model parameters: Column density N$_H$, in units of 10$^{22}$~cm$^{-2}$; covering factor C$_F$; slope $\Gamma$ of the continuum power law;
ratio F/F$_*$ between the absorption-corrected flux F and that measured in the third orbit (INT~4). The continuum reflection and the emission lines are fixed to the values shown in Table~1.
}}
\end{table}
  
\subsection{Time-resolved analysis of the occultations.}
{ The previous step of the analysis shows that it is possible to interpret the spectral variations in the first and second orbit as entirely due to occultations by clouds crossing the line of sight to the X-ray source. Here we assume that this is the correct interpretation, and that no other continuum components are contributing to the {\em spectral} variability (while we allow for possible intrinsic flux changes). In this way, we are able to perform a more detailed spectral analysis on shorter time scales, requiring a constant shape of the continuum components.} 
We divided the first two orbits in several short intervals, as shown in Fig.~3 and~4, and performed a complete spectral analysis of each interval. 
\begin{figure}
\includegraphics[width=8.0cm,angle=0]{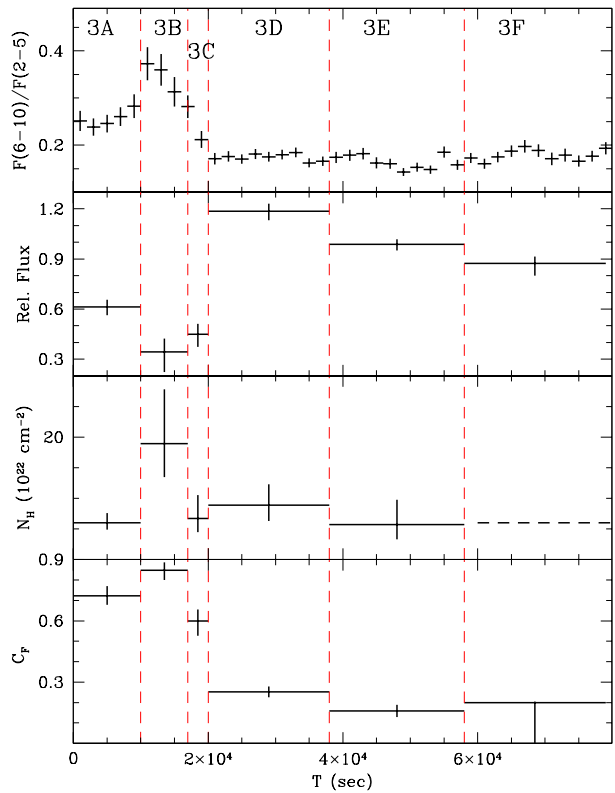}
\caption{Same as Fig.~3, for the second orbit of the {\em XMM-Newton} observation (INT~3 in Fig.~1). 
}
\label{totfit}
\end{figure}
\begin{table}
\caption{Time resolved spectral analysis}
\centerline{\begin{tabular}{lcccc}
\hline
INT   &  N$_H$              & C$_F$                    & F/F$_*$                 & $\chi^2$/d.o.f.\\
\hline
1A    &24.8$_{-5.1}^{+6.8}$ & 0.89$_{- 0.02}^{+ 0.02}$ & 0.57$_{- 0.09}^{+0.12}$ & 382/434 \\
1B    &15.5$_{-3.1}^{+4.5}$ & 0.89$_{- 0.04}^{+ 0.03}$ & 0.43$_{- 0.05}^{+0.07}$ & 389/429 \\
1C    & 9.1$_{-1.3}^{+1.6}$ & 0.82$_{- 0.03}^{+ 0.03}$ & 0.52$_{- 0.03}^{+0.04}$ & 525/595 \\
1D    &10.5$_{-1.5}^{+1.7}$ & 0.71$_{- 0.02}^{+ 0.02}$ & 0.92$_{- 0.05}^{+0.06}$ & 827/763 \\
1E    & 5.4$_{-1.0}^{+1.5}$ & 0.68$_{- 0.05}^{+ 0.05}$ & 0.74$_{- 0.04}^{+0.05}$ & 460/518 \\
\hline
2A    & 7.9$_{-1.5}^{+2.0}$ & 0.63$_{- 0.03}^{+ 0.03}$ & 1.01$_{- 0.06}^{+0.07}$ & 607/591 \\
2B    &10.8$_{-2.1}^{+2.8}$ & 0.69$_{- 0.03}^{+ 0.03}$ & 0.92$_{- 0.07}^{+0.08}$ & 571/528 \\
2C    &14.6$_{-4.7}^{+7.3}$ & 0.75$_{- 0.05}^{+ 0.04}$ & 0.58$_{- 0.08}^{+0.12}$ & 365/367 \\
2D    &17.1$_{-3.9}^{+6.0}$ & 0.84$_{- 0.03}^{+ 0.03}$ & 0.57$_{- 0.07}^{+0.10}$ & 372/347 \\
2E    &18.4$_{-3.3}^{+3.9}$ & 0.97$_{- 0.05}^{+ 0.02}$ & 0.52$_{- 0.07}^{+0.09}$ & 285/288 \\
\hline
3A    & 6.0$^{+1.5}_{-1.1}$  & 0.72$^{+0.05}_{-0.04}$ & 0.61$^{+0.05}_{-0.04}$ & 559/572 \\
3B    & 18.9$^{+8.9}_{-5.5}$ & 0.85$^{+0.04}_{-0.05}$ & 0.34$^{+0.12}_{-0.08}$ & 355/340 \\
3C    & 6.7$^{+3.8}_{-2.2}$  & 0.60$^{+0.06}_{-0.07}$ & 0.45$^{+0.08}_{-0.06}$ & 284/268  \\
3D    & 8.9$^{+3.4}_{-2.6}$  & 0.25$^{+0.03}_{-0.03}$ & 1.19$^{+0.05}_{-0.05}$ & 1279/1234 \\
3E    & 5.7$^{+4.0}_{-2.4}$  & 0.16$^{+0.03}_{-0.04}$ & 0.99$^{+0.04}_{-0.03}$ & 1295/1227 \\
3F    & 6$^a$                & $<$0.18                & 0.87$^{+0.07}_{-0.04}$ & 1188/1174 \\
\hline
\end{tabular}}
\footnotesize{Model parameters: Column density N$_H$, in units of 10$^{22}$~cm$^{-2}$; covering factor C$_F$,
ratio F/F$_*$ between the absorption-corrected flux F and that measured in the third orbit (INT~4). { All the other parameters are fixed to the best fit vaues of INT~4, listed in Table~1.} $^a$: the column density is
fixed, due to the complete degeneracy with the covering factor when the latter is compatible with C$_F$=0. }
\end{table}

The length of the intervals is chosen in order to have a balance between the need to
sample the changes in hardness ratio in short time scales, and the need to have enough counts in each spectrum to significantly constrain the absorber parameters. 
{ The continuum 
spectral parameters (spectral slope, reflection component) are those obtained in the fit to the third and fourth orbit discussed in the previous step of the analysis, and listed in Table~1. The continuum flux is free to vary.}

The main results of this analysis are shown in the lower panels of Fig.~3 and 4, and in Table~2, and can be summarized as follows:\\
-1. In all cases, we obtain a good spectral fit. This implies that the observed 
spectral variations { can be reproduced satisfactorily by} absorbing clouds moving across the line of sight to the X-ray source.\\
-2. The absorbing clouds have column densities in the range 1-3$\times$10$^{23}$~cm$^{-2}$, while the covering factors vary from zero up to $\sim$90\% at the peak of the eclipses.\\
-3. We interpret the variations observed in the first orbit as due to two different clouds, one uncovering the source during intervals 1A to 1E, the other covering the source during intervals 2A-2E. 
Since the minimum measured covering factor is $\sim$60\% (Table~2), the two eclipses overlap in time, in a way that cannot be uniquely determined. The bottom panel of Fig.~3 shows two possible evolutions of the single occultations, both resulting in the observed light curve. \\
-4. The two iron absorption lines have been included in all fits. 
In principle, this may be a powerful way to investigate possible variations of the ionized absorber. However, the
signal-to-noise of the single intervals is not enough for a detailed analysis: in all the intervals 
we find line parameters in agreement (within the errors) with the hypothesis of no variations apart from the outflow velocity (Table~4).
The significance of these detections varies considerably among the different intervals: in some cases both lines are
detected at a $>$4~$\sigma$ statistical significance, in others we can only estimate upper limits. However, this 
appears to be related to the total counts available in the continuum at the line energies, rather than to variations
in the lines' strength. As a consequence, the upper limits in the intervals with no detections are of the same order of 
(or even slightly larger than) the values measured in the intervals with a high continuum level.
For this reason, we will not discuss possible variations of the absorption lines further. { Let us stress that unconstraining upper limits are found only in some sub-intervals of INT~1-3; as we said before, the analysis of INT~1-4 in wider time bins is instead very constraining.}

-5. The results listed above do not depend significantly on how we fit the variability of the other spectral components.
In particular, since our model consists of three main components (the direct power law emission, the cold reflection, and the iron emission lines), we can assume different scenarios: 1) the continuum reflection {\em and} the emission lines (both the narrow and the broad one)
originate from the accretion disc, close to the primary X-ray source, and follow the continuum variations with a time lag much shorter than our time bins; 
2) the reflection components (continuum and narrow line) are emitted far from the primary source, and are constant during the whole observation, while the broad emission line follows the continuum variations; 
3) the continuum reflection and both emission lines remain constant.
We repeated our analysis in these three scenarios, and found that the results on the continuum occultations and on the absorption lines hold with the same significance in all cases. Therefore, we do not discuss this issue further. We only note, for completeness, that statistically the first and second scenario are equally acceptable, while the 
third scenario produces significantly worse fits, in particular with strong residuals at the broad iron line energies.
\begin{table}
\caption{Absorption line parameters}
\centerline{\begin{tabular}{lcccc}
\hline
INT     & v$^a$                    & EW$_1^b$          & EW$_2^b$          & $\Delta\chi^2$ \\
\hline
1  &  11000$^{+900}_{-1100}$    & 61$^{+15}_{-14}$  & 50$^{+18}_{-18}$  & 70 \\
2  &  3000$^{+1200}_{-1200}$    & 43$^{+18}_{-17}$  & 64$^{+18}_{-17}$  & 43 \\
3A-C& 16000$^{+3000}_{-3000}$   & 58$^{+26}_{-40}$  & 46$^{+30}_{-35}$  & 20 \\        
3D & 8100$^{+3500}_{-3600}$     & 39$^{+22}_{-21}$  & 35$^{+20}_{-20}$  &  6 \\
3E-F & $<13000$                 & $<30$             & $<30$             &  4 \\
4  & $<10000$                   & $<18$             & $<20$             &  3 \\
\hline
\hline
\end{tabular}}
\footnotesize{Model parameters for the absorption lines, in the three intervals discussed in the text.
No line is detected in the fourth interval (corresponding to the third and fourth orbit, Fig.~1).
$^a$: velocity in km/s; $^b$: equivalent width in eV of the Fe~XXV K$\alpha$ and K$\beta$ lines, respectively.}
\end{table}

\begin{figure}
\includegraphics[width=8.5cm,angle=0]{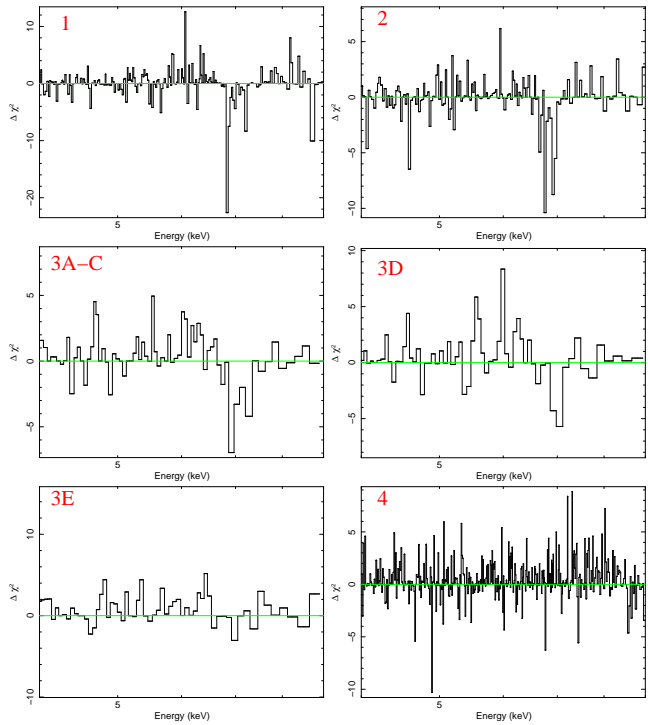}
\caption{$\chi^2$ residuals for the six intervals considered in Table~3, corresponding to the three occultation events (1, 2 and 3A-C), the first interval after the third occultation (3D), the remainder of the second orbit  (3E-F), and
the whole third and fourth orbits (4).
}
\label{totfit}
\end{figure}

\section{Discussion}

{ We presented a time-resolved spectral analysis of a long ($\sim$700~ks) {\em XMM-Newton} observation of 
the NLS1 galaxy Mrk~766, suggesting the occurrence of occultations of the X-ray source due 
to circumnuclear clouds.}

While similar events have been seen in obscured AGNs (e.g. Wang et al.~2010, Risaliti et al.~2010, Maiolino et al.~2010), our analysis demonstrates that such events can also be observed also in type~1 sources, which are normally 
unobscured. Indeed, the spectrum of Mrk~766 shows on average little obscuration: the eclipses occur in a 
small fraction ($\sim$20\%) of the observing time, and would be easily missed in a global spectral analysis.

Interestingly, the analysis of T07, where the spectra are extracted from fixed 25~ks intervals, provides
indications of the occultations described here: the N$_H$ values of the cold absorption component 
reported there are on average marginally significant, but show local peaks and higher statistical significance 
in the time
intervals where we find the occultation events. The only difference with our analysis is given by the 
choice of the time intervals, which in our work are optimized based on Fig.~1.

In the following sections we discuss (1) the size and distance of the obscuring clouds, as inferred from the observed occultations; (2) the internal structure of the clouds, as suggested by the temporal evolution of the occultations;
(3) a possible general structure of the circumnuclear medium of Mrk~766 accounting for all the results discussed in
this paper. 

\subsection{Size and location of the obscuring clouds}

The spectral and temporal parameters of the three occultations provide information on the size and
distance of the clouds from the X-ray source. 
Assuming that the X-ray source has a linear size of at least 5 gravitational radii, R$_G$, the velocity
of the obscuring cloud should be v$>$2.5$\times$10$^{3}$ M$_6$~T$_4^{-1}$~$\sqrt{\Delta C_F}$~km~s$^{-1}$, where M$_6$ is the black hole mass in units of 10$^6$~M$_\odot$, T$_4$ is the occultation time in units of 10$^4$~s, and $\Delta$C$_F$ is
the covering factor variation during the eclipse.
The black hole mass estimate for Mrk~766, obtained from reverberation mapping of H$\beta$, is M$_6$=1.8$\pm$1.5 
(Bentz et al.~2009). 
The durations of the occultations seen in the first orbit are of the order of 40~ks, with a 
covering factor changing by $\sim$40\%. This implies v$_{1,2}$$>$800~km~s$^{-1}$. We note that this value is not precisely estimated, because the end of the fist occultation probably overlaps with the beginning of the second one. We will discuss this issue in the next Subsection. However our conclusions are not affected by this relatively small uncertainty. For the third occultation we have instead T$\sim10$~ks and
$\Delta$C$_F$$\sim$80\%, which gives v$_3$$>$4,000~km~s$^{-1}$. We note that the ratio between the velocities depends only on the ratio between the occultation times (and on the assumption of an axisymmetric cloud shape with respect to the line of sight). We have therefore v$_3$$\sim$5v$_{1,2}$ regardless of the
uncertainties on the size of the X-ray source.
The density and distance of the clouds  from the central black hole 
can be easily derived from the above estimates(assuming Keplerian velocity). We obtain a density of the order of 10$^{10}$-10$^{11}$~cm$^{-3}$,
and a distance of 10$^3$-10$^4$~R$_G$. 

This simple argument suggests that the X-ray eclipsing clouds have 
the physical properties and location of the BLR clouds.

{ We note that the above considerations can provide an estimate of the size of the X-ray source. The assumption made above of a size of 5~R$_G$ can be considered as a ``physical'' lower limit, useful to estimate lower limits for the cloud velocities. Since in the above scheme these velocities are linearly dependent on the source size, we can reverse the argument, and estimate an upper limit of the size of the X-ray source based on the highest physically acceptable cloud velocity. 
Considering that the third cloud has an outflow velocity of $\sim$16,000~km~s$^{-1}$, we can assume a velocity on the plane of the sky of the same order. This implies a linear size of the X-ray source of 20~R$_G$. Adopting instead as a ``hard upper limit'' the highest outflowing velocity detected in low-luminosity AGNs (v/c$\sim$0.3, Tombesi et al.~2010), we would obtain a size of $\sim$100~R$_G$. Summarizing, the duration of the occultations are a direct observational probe of the compactness of the X-ray source, with a best estimate below $\sim$20~R$_G$, and a maximum size of the order of 100~R$_G$.}
 
\subsection{Internal structure of the obscuring clouds}
Hints on the internal structure of the eclipsing clouds are obtained from the temporal evolution of the
model parameters, as listed in Table~2 , 3 and 4.
The clearest indication that the clouds have a complex structure, with strong ionization gradients, is
provided by the coincidence between the iron absorption lines and the cold absorption: while the presence of the
lines is clear during all the three eclipses, no such feature is present in any other time interval, with
high statistical confidence (Table~4). Moreover, the outflow velocities measured in the three occultations are
not compatible with each other. This is relevant in two respects:
1) it demonstrates that the three occultations are due to three different clouds; 
2) it reveals a large range in outflow velocities
among the obscuring clouds, comparable to the range spanned by the transverse velocities. 
We discuss in detail these two aspects, and the column density structure of the eclipsing clouds.

{\bf 1) The three eclipsing clouds.} 
While it was obvious from the previous analysis that the third occultation is due to a single cloud isolated from
the other two, the interpretation of the occultations in the first orbit was less straightforward. In particular, the
lowest measured covering factor in the first orbit is about 0.6. This means that either two different clouds partly overlap at the same time, or that a single cloud with a complex structure (a hole, or a concave shape) is
transiting across the line of sight. The association of two different outflowing velocities to the two phases
of the eclipse seen in the first orbit (INT~1 and INT~2) strongly suggests that the former scenario is the correct one.  

If two different clouds are covering the source in the first orbit, the slopes observed in Fig.~3 are not directly related to the clouds velocites along the plane of the sky. Several combinations of the motion of the individual clouds can reproduce the observed pattern. In Fig.~3 we show the two extreme configurations acceptable in the case of a constant slope of the CF time evolution. 
In both cases, the two clouds are simultaneously present along the line of sight in one of the two phases. However, we point out that these are just two examples of many possible
combinations: there is no strong physical reason to have a constant slope in the covering factor curves
(the geometrical shape of the cloud and of the X-ray source can easily alter this, as shown, for example, in the case
of NGC~1365, Maiolino et al.~2010), and if we relax the constraint on the constant slope, there are infinite ways to reproduce the observed behaviour. In principle, this degeneracy may be solved through the observation of the absorption lines:
the presence of both clouds along the line of sight would imply the detection of both line pairs. However,
this check is not decisive with the available S/N: we tried to add a second pair of absorption lines in INT 1, with
a blueshift velocity of 3000~km~s$^{-1}$ (as measured in INT 2), and in INT~2 with a blueshift of 11,000~km~s$^{-1}$ (as
measured in INT~1), and in both cases the additional lines are not detected, with upper limits on their equivalent widths of the
order of 30~eV.\footnote{We note that, adding to the complication of the analysis, the difference between the velocities of the two clouds, $\Delta$v$\sim$8,000~km~s$^{-1}$, corresponds almost exactly to the difference in frequency between 
FeXXV~K$\alpha$ anf FeXXVI~K$\alpha$. As a consequence, a contribution from both clouds to just one of the observed lines is possible.} We also repeated this check using only intervals 1E and 2A (where the contribution of both clouds may
be stronger) and obtained an upper limit of the order of 80~eV, i.e. higher than the typical values in Table~4. We conclude that the data suggest that on average in both INT~1 and INT~2  one single cloud is dominant (at least for the
highly ionized component), but we cannot exclude a significant overlap in the central part of the first orbit.

The conclusions of this part of the analysis are that (1) three separate clouds are responsible of the observed occultations,
each with a different velocity along the plane of the sky, and along the line of sight; and that (2) it is not possible
to uniquely disentangle the path of the two individual clouds responsible for the two occultations observed in the first orbit.

{\bf 2) Ionization structure}.  The lines detected during the occultations are due to hydrogen-like and helium-like iron ions.
The physical properties (density, column density, ionization state)
 of the gas responsible for these absorptions are not easily constrained with the observed lines.
The measured parameters (equivalent widths and line flux ratios) can be obtained with several combinations of column density, ionization parameter, and turbulent velocity of the gas. Morever, the equivalent widths scale linearly with the (unknown) iron abundance. A general treatment of this complex problem is presented in Bianchi et al.~(2005). Here we only stress the two conclusions 
that can be safely drawn from the observed data: the ionization parameter of the gas must be quite high ($\xi$$>$1,000 erg~cm~s$^{-1}$) and
the gas column density must be higher than a few 10$^{22}$~cm$^{-2}$. On the other hand, the absorption observed
in the continuum spectrum is due to nearly neutral material. We tried to replace the neutral absorption in our models with a 
ionized one, with a free ionization parameter, $\xi$ and we obtained again a best fit with a neutral absorber, and a limit on
the ionization $\xi$$<$30~erg~cm~s$^{-1}$. 
These two phases can be explained as due to a "cometary" structure, with a low-density, high ionization "tail" originating from the high-density cloud head. The tail may be either behind the head, with respect to the observer,
or in front of the head, as in a "real" solar system comet (Collin-Souffrin et al.~1988). 
In the former case, the whole cloud, including the dense core, must be in outflow, and the tail may be due to friction with 
a lower density, higher temperature gas phase surrounding the cloud. Instead, in the scenario where the tail is in front of the head, the dense core can be in a purely Keplerian motion, and the outflowing tail may be due to radiation pressure.
While we cannot distinguish between these two scenarios based on the X-ray spectra only, we can safely rule out a  spherically
symmetric cloud structure, without a ``head'' and a ``tail''. This is due to the requirements of (1) having both 
the hot and cold phases along the line of sight at the same time, and (2) 
a difference between the ionization parameters of the two
phases of at least a factor of $\sim$40. It is impossible to satisfy both these conditions with a spherical cloud: the low-density external part, in order to have the required column density of at least a few 10$^{22}$~cm$^{-2}$, must have a linear size of at least 10~times that of the cold, dense part. In a spherical geometry, this would imply the presence of high ionization absorption features for a long time before and after the observed eclipses.  
Regarding the density structure of the clouds,
we may have either a sudden density drop, or a continuous density gradient. In the latter case, intermediate ionization
states would be present in the cloud. Both these scenarios are possible, and our observations cannot distinguish among them,
because the tail lies along the same line of sight as the neutral head. This hides the effects of 
possible ionization gradients, since the cold absorption removes all the marks of a warm absorber, except for the
iron-K absorption lines, which are the only strong features at energies higher than the photoelectric cut~off.
We note that a low-column density tail not covered by the neutral head may be present in the third cloud. A low-significance
indication of such a component may be the marginal detection of the absorpion lines in the interval 3D (Fig.~5, Table~3), i.e. after the neutral
eclipse is over. This line would not be statistically significant if detected in an isolated spectrum ($\Delta\chi^2$=6, Table~3)
however, given the presence of similar lines in the previous spectra, it may be possible that this is a real feature.
A similar analysis is not possible for the first two clouds, given the overlap between the two eclipses, and the lack of monitoring during the final phase of the second eclipse. 
\begin{figure}
\centerline{
\includegraphics[width=8.5cm,angle=0]{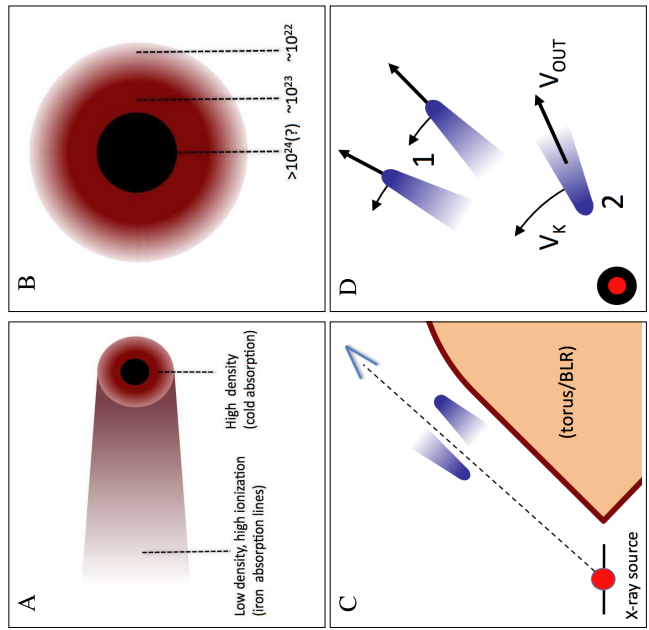}}
\caption{Scheme of the proposed structure of the absorber responsible for the observed occultations. Panel~1: structure of an eclipsing cloud, as viewed from a direction orthogonal to the line of sight. Panel~2: same, as viewed from the line of sight, with the 
ionized tail covered by the neutral head (either behind or in front of it). Panel~3: a possible structure of the circumnuclear gas in Mrk~766, as viewed from the disc/torus plane. Panel~4: velocity components of the eclipsing clouds, as viewed from the disc/torus axis. The cloud tail can be either in front (cloud 2) or behind (1) the cold core with respect to the source.}
\end{figure}

{\bf 3) Column density structure of the eclipsing clouds.} Hints on the column density (N$_H$) gradients across the line of sight can be inferred by comparing the N$_H$ and covering factor (C$_F$) light curves in Fig.~3 and~4. 
In general, the evidence of N$_H$ changes is not strong (the case of constant column density can be rejected at a $\sim$2$\sigma$ level for INT~2, and at a $\sim$3$\sigma$ level fot INT~1 and INT~3). However, we clearly see a general trend of a column density following the covering factor variations: the highest C$_F$ values are found together with the highest N$_H$, and vice-versa.
This is especially clear in INT~3 (Fig.~4) where we see all the phases of the eclipse. We interpret this as an evidence of a
column denisty gradient across the line of sight, as shown in Fig.~6 (Panel~2). { The possible relation between $C_F$ and $N_H$ may be due to the partial degeneracy between these two parameters in the spectral fits. In order to check this, and in general to verify the precision of our error estimates, we calculated the $C_F$-N$_H$ contour plots for several representative intervals of the first and second orbit. The results, shown in Fig.~7, demonstrate that the degeneracy between the two parameters is not severe, and that the single-parameter errors used in this work are representative of the real  uncertainties. }
\begin{figure}
\includegraphics[width=8.5cm,angle=0]{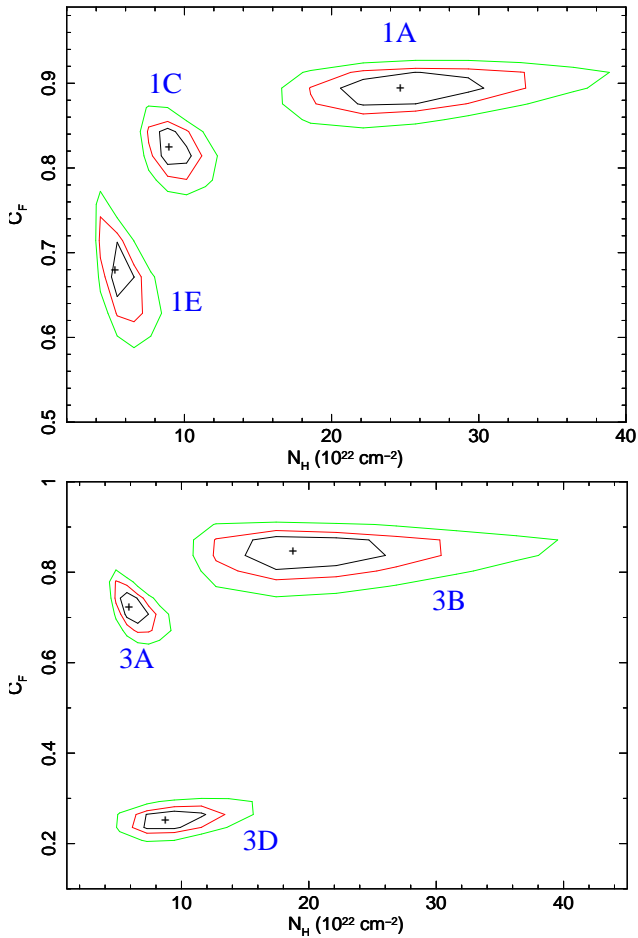}
\caption{Covering factor (C$_F$) - Column density (N$_H$) contour plots for several representative intervals.
Labels refer to the intervals as defined in Fig.~3 and~4.
}
\end{figure}

At a more speculative level, we notice that the intrinsic continuum flux during the third eclipse is anticorrelated with 
$N_H$ and C$_F$. This may be a simple coincidence, given the strong and fast intrinsic variability of this source (Fig.~1), or
otherwise it could be the indication of the presence of a Compton-thick core in the cloud. The effect of this component would be
to decrease the observed flux, with no significant change in the spectral shape (except for a small flattening
of the total spectrum, due to the higher fraction of reflected continuum). The possible Compton-thick core is also indicated in
Fig.~6 with a question mark. 
 
\subsection{A general picture of the circumnuclear medium}

The above analysis shows that, { within the adopted scenario,} we can infer several physical and geometrical properties of the eclipsing clouds.
Here we discuss a general view of the circumnuclear region of Mrk~766 accounting for the observed occultations.

The main point that we want to stress is that the observed clouds are {\em not} expected to be "average" Broad Line Region clouds.
Mrk~766 is a well known Narrow-Line Seyfert 1, with optical line widths of the order of 1,100~km~s$^{-1}$ (Grupe et al.~2004) centered 
at the rest-frame line energies,  and on average with no absorption along the line of sight. This is also the case for the
observation discussed here: the source remains completely unobscured for $\sim$80\% of the time. 
The scenario we propose is described in Fig.~6: the "average" broad line clouds form a toroidal structure around the
central source. The clouds responsible of our occultations are out of the main distribution, probably driven by radiation pressure from
the central accretion disc. The proposed structure is in agreement with wind-torus models such as in Elvis~(2000), Proga et al.~(2000) or
Risaliti \& Elvis~(2009), where a small part of 
the circumnuclear gas is accelerated at outflowing velocities of the order of those observed here, 
in a nearly radial direction, while most of the gas falls down into the disk, forming a toroidal stucture. 
The simulations in Proga et al.~(2000) show a large range in density (and therefore in ionization) inside the wind region.
Since in Mrk~766 we are not observing a constant wind, it is possible that the observed clouds are "outliers" slightly scattered out with respect to the main wind stream. 

The clouds detected in Mrk~766 are completely different from the ones analyzed in a recent {\em Suzaku} observation of NGC~1365 (Maiolino et al.~2010), where we were also able to observe comet-shaped clouds: in the case of NGC~1365 the eclipses are much more common, the motion of the clouds is likely dominated by the radial component, and the tails lie on the plane of sky. NGC~1365 is a heavily obscured, optically type~1.8 source, so we believe that the line of sight to the X-ray source is crossed by ‘‘standard'' clouds lying in the bulk of the toroidal Broad Line Region.


\section{Conclusions}

We have presented a time-resolved spectral analysis of a long {\em XMM-Newton} observation of the NLS1 galaxy Mrk~766,
suggesting the presence of three occultation events due to three different clouds crossing the line of sight.
{ Within this scheme,} the spectral analysis of the single intervals revealed important features of the absorber structure:\\
1) The obscuring clouds have  neautral cores with column densities of a few 10$^{23}$~cm$^{-2}$, density of 10$^{10}-10^{11}$~cm$^{-3}$, and linear dimensions of the order of 10$^{12}$-10$^{13}$~cm. The X-ray source has a similar size, 
corresponding to a few gravitational radii for the estimated black hole mass of Mrk~766 of 1.8$\times$10$^6$~M$_\odot$.\\
2) The detection of Fe~XXV and Fe~XXVI absorption lines during all the three occultations 
suggests the presence of highly ionized cloud tails, outflowing with velocities from 3,000 to 15,000~km~s$^{-1}$. \\

Our analysis shows that occultation events, already observed in several obscured AGN, can happen also in type 1 objects,
which are on average unobscured. When this happens, allowing for possible absorption variability is crucial both to
obtain information on the structure of the X-ray absorber (and of the broad emission line clouds) and to perform a correct
estimate of the parameters of the X-ray spectrum. 

We are currently expanding our time-resolved analysis to several tens of 
long archival observations of bright AGNs, with the aim of
increasing the statistics on X-ray occultation events, and building a well defined sample in order to  estimate the 
frequency of such ``black hole eclipses''.

\section*{Acknowledgements}
This work has been funded by NASA Grant NNX07AR90G. 



\begin{thebibliography}{}
\bibitem[\protect\citeauthoryear{Bentz et al.}{2009}]{2009ApJ...705..199B} 
Bentz M.~C., et al., 2009, ApJ, 705, 199
\bibitem[\protect\citeauthoryear{Bianchi et 
al.}{2005}]{2005MNRAS.357..599B} Bianchi S., Matt G., Nicastro F., Porquet 
D., Dubau J., 2005, MNRAS, 357, 599
\bibitem[\protect\citeauthoryear{Boller et 
al.}{2001}]{2001A&A...365L.146B} Boller T., Keil R., Tr{\"u}mper J., O'Brien P.~T., Reeves J., Page M., 2001, A\&A, 365, L146 
\bibitem[\protect\citeauthoryear{Collin-Souffrin et 
al.}{1988}]{1988MNRAS.232..539C} Collin-Souffrin S., Dyson J.~E., McDowell 
J.~C., Perry J.~J., 1988, MNRAS, 232, 539 
\bibitem[\protect\citeauthoryear{Elvis}{2000}]{2000ApJ...545...63E} Elvis 
M., 2000, ApJ, 545, 63 
\bibitem[\protect\citeauthoryear{George 
\& Fabian}{1991}]{1991MNRAS.249..352G} George I.~M., Fabian A.~C., 1991, MNRAS, 249, 352 
\bibitem[\protect\citeauthoryear{Grupe et al.}{2004}]{2004AJ....127..156G} 
Grupe D., Wills B.~J., Leighly K.~M., Meusinger H., 2004, AJ, 127, 156 
\bibitem[\protect\citeauthoryear{Leighly et 
al.}{1996}]{1996ApJ...469..147L} Leighly K.~M., Mushotzky R.~F., Yaqoob T., 
Kunieda H., Edelson R., 1996, ApJ, 469, 147
\bibitem[\protect\citeauthoryear{Magdziarz 
\& Zdziarski}{1995}]{1995MNRAS.273..837M} Magdziarz P., Zdziarski A.~A., 1995, MNRAS, 273, 837
\bibitem[\protect\citeauthoryear{Maiolino et 
al.}{2010}]{2010arXiv1005.3365M} Maiolino R., et al., 2010, arXiv, 
arXiv:1005.3365
\bibitem[\protect\citeauthoryear{Matt et 
al.}{2000}]{2000A&A...363..863M} Matt G., Perola G.~C., Fiore F., Guainazzi M., Nicastro F., Piro L., 2000, A\&A, 363, 863 
\bibitem[\protect\citeauthoryear{McHardy et 
al.}{2004}]{2004MNRAS.348..783M} McHardy I.~M., Papadakis I.~E., Uttley P., 
Page M.~J., Mason K.~O., 2004, MNRAS, 348, 783
\bibitem[\protect\citeauthoryear{Miller et 
al.}{2007}]{2007A&A...463..131M} Miller L., Turner T.~J., Reeves J.~N., George I.~M., Kraemer S.~B., Wingert B., 2007, A\&A, 463, 131 
\bibitem[\protect\citeauthoryear{Molendi 
\& Maccacaro}{1994}]{1994A&A...291..420M} Molendi S., Maccacaro T., 1994, A\&A, 291, 420 
\bibitem[\protect\citeauthoryear{Nandra et al.}{1997}]{1997ApJ...476...70N} 
Nandra K., George I.~M., Mushotzky R.~F., Turner T.~J., Yaqoob T., 1997, 
ApJ, 476, 70
\bibitem[\protect\citeauthoryear{Page et 
al.}{2001}]{2001A&A...365L.152P} Page M.~J., et al., 2001, A\&A, 365, L152
\bibitem[\protect\citeauthoryear{Pounds et al.}{2003}]{2003MNRAS.342.1147P} 
Pounds K.~A., Reeves J.~N., Page K.~L., Wynn G.~A., O'Brien P.~T., 2003, 
MNRAS, 342, 1147 
\bibitem[\protect\citeauthoryear{Proga, Stone, 
\& Kallman}{2000}]{2000ApJ...543..686P} Proga D., Stone J.~M., Kallman T.~R., 2000, ApJ, 543, 686
\bibitem[\protect\citeauthoryear{Risaliti et 
al.}{2005}]{2005ApJ...623L..93R} Risaliti G., Elvis M., Fabbiano G., Baldi 
A., Zezas A., 2005A, ApJ, 623, L93 (R05A)
\bibitem[\protect\citeauthoryear{Risaliti et 
al.}{2007}]{2007ApJ...659L.111R} Risaliti G., Elvis M., Fabbiano G., Baldi 
A., Zezas A., Salvati M., 2007, ApJ, 659, L111 (R07) 
\bibitem[\protect\citeauthoryear{Risaliti et 
al.}{2009}]{2009MNRAS.393L...1R} Risaliti G., et al., 2009A, MNRAS, 393, L1 
\bibitem[\protect\citeauthoryear{Risaliti et 
al.}{2009}]{2009ApJ...696..160R} Risaliti G., et al., 2009B, ApJ, 696, 160 
\bibitem[\protect\citeauthoryear{Risaliti 
\& Elvis}{2009}]{2009arXiv0911.0958R} Risaliti G., Elvis M., 2009, arXiv, arXiv:0911.0958
\bibitem[\protect\citeauthoryear{Risaliti et 
al.}{2010}]{2010MNRAS.406L..20R} Risaliti G., Elvis M., Bianchi S., Matt 
G., 2010, MNRAS, 406, L20
\bibitem[\protect\citeauthoryear{Ross 
\& Fabian}{2005}]{2005MNRAS.358..211R} Ross R.~R., Fabian A.~C., 2005, MNRAS, 358, 211 
\bibitem[\protect\citeauthoryear{Turner et 
al.}{2007}]{2007A&A...475..121T} Turner T.~J., Miller L., Reeves J.~N., Kraemer S.~B., 2007, A\&A, 475, 121 (T07)
\bibitem[\protect\citeauthoryear{Vaughan 
\& Fabian}{2003}]{2003MNRAS.341..496V} Vaughan S., Fabian A.~C., 2003, MNRAS, 341, 496 
\bibitem[\protect\citeauthoryear{Wang et al.}{2010}]{2010ApJ...714.1497W} 
Wang J., Risaliti G., Fabbiano G., Elvis M., Zezas A., Karovska M., 2010, 
ApJ, 714, 1497






 




\end{thebibliography}
\end{document}